\documentclass[%
 aip,
 sd,%
 amsmath,amssymb,
 reprint,%
]{revtex4-1}

\usepackage[utf8]{inputenc}

\usepackage{graphicx}
\usepackage{dcolumn}
\usepackage{bm}
\usepackage{todonotes}
\usepackage{upgreek}

\begin{document}


\title{Green’s function-based control-oriented modeling of electric field for dielectrophoresis}

\affiliation{ 
Faculty of Electrical Engineering, Czech Technical University in Prague, Czech Republic.
}%

\author{Martin Gurtner}
 \email{martin.gurtner@fel.cvut.cz}
\author{Kristian Hengster-Movric}%

\author{Zdeněk Hurák}
 \homepage{http://aa4cc.dce.fel.cvut.cz.}

\date{\today}

\begin{abstract}
In this paper, we propose a novel approach to obtaining a reliable and simple mathematical model of a dielectrophoretic force for model-based feedback micromanipulation. Any such model is expected to sufficiently accurately relate the voltages (electric potentials) applied to the electrodes to the resulting forces exerted on microparticles at given locations in the workspace. This model also has to be computationally simple enough to be used in real time as required by model-based feedback control. Most existing models involve solving two- or three-dimensional mixed boundary value problems. As such, they are usually analytically intractable and have to be solved numerically instead. A numerical solution is, however, infeasible in real time, hence such models are not suitable for feedback control. We present a novel approximation of the boundary value data for which a closed-form analytical solution is feasible; we solve a mixed boundary value problem numerically off-line only once, and based on this solution we approximate the mixed boundary conditions by Dirichlet boundary conditions. This way we get an approximated boundary value problem allowing the application of the analytical framework of Green's functions. Thus obtained closed-form analytical solution is amenable to real-time use and closely matches the numerical solution of the original exact problem.
%
\end{abstract}

\maketitle


\section{Introduction} 
\label{sec:introduction}
Since its invention by H. Pohl in the 1950s and 1960s\cite{Pohl1958Some,Pohl1966Separation}, \textit{dielectrophoresis (DEP)} has proved an efficient tool for transportation, separation, and characterization of microparticles such as e.g. biological cells (see Refs.~\onlinecite{Pethig2010Review,Jones2003Basic} for a recent survey and comprehensive introduction). More often than not, DEP is used to manipulate ensembles of large numbers of microparticles. However, recently some attempts were successful to use DEP in a feedback control scheme for a high accuracy noncontact manipulation of a single microparticle\cite{Zemanek2015Feedback,Kharboutly2013High,Kharboutly20122D,Edwards2012Electric}; these developments can be viewed as a reopening of the topic first started in the 1990s\cite{Kaler1990Dielectrophoretic}. The technology has also boosted development in this area; there are reported CMOS chips integrating both actuation and sensing and thus enabling individual and independent manipulation of thousands of cells\cite{Manaresi2003Cmos}. This technology has later been commercialized by Silicon Biosystems as a commercial product called DEPArray\textsuperscript{TM}. These non-contact tweezers are usually based on a feedback control scheme, typically invoking an automatic visual tracking. The feedback control, in turn, requires a sufficiently accurate mathematical model of the underlying physical phenomenon of DEP. The relationship between the voltages applied to the microelectrodes and the DEP force exerted on a microparticle located at a given position needs to be evaluated periodically as the microparticle moves around the workspace. Sampling periods on a time-scale of few tens of milliseconds or even a few milliseconds are not unusual. The commonly used approaches to modeling DEP---which are based on numerical solution of the corresponding \textit{Boundary Value Problem (BVP)}, typically using \textit{Finite Elements Method (FEM)} or \textit{Method of Moments (MOM)}---are not feasible in real time. It is possible to precompute and store these solutions in a computer memory (as reported in Ref. \onlinecite{Zemanek2015Feedback}) but this approach imposes stringent requirements on the volume of data stored. There are approaches described in the literature that provide analytical solutions\cite{Morgan2001Dielectrophoretic,Chang2003ClosedForm,Wang1996Theoretical}; however, they are only usable for simple electrode arrays and fixed harmonic voltage signals applied to the electrodes while the feedback control requires the ability to change the voltages in real time.

In this paper, we propose a modeling methodology that provides a computationally simple yet sufficiently accurate model of a DEP force for the purposes of feedback micromanipulation. We propose to combine numerical and analytical approaches to modeling of DEP. Existing models of DEP usually involve a numerical solution of an analytically intractable \textit{mixed Boundary Value Problem (mBVP)} for the potential in the workspace. As the numerical solution is infeasible in real time and might be too large for storing in a computer memory, it is desirable to find a closed-form approximate analytical expression for the potential. To find such an expression, we solve numerically the original mBVP. Based on this numerical solution, we approximate the mBVP by a BVP for which the closed-form solution can be found by Green's functions. Using the approximate closed-form expression for the potential we obtain a model of DEP force that is computationally effective and requires almost no memory space. The numerical solution of the mBVP needs to be computed off-line and only once. Thus the high computational burden associated with the numerical solution is carried out off-line and the feedback control system uses only the approximate closed-form analytical solution in real time.



The paper is organized as follows. In Section~\ref{sec:feedback_manipulation_by_dielectrophoresis}, we briefly present the commonly used dipole model of DEP and show what prevents its direct use in feedback micromanipulation. In Section~\ref{sec:derivation_and_results} we propose a control-oriented model derived from the dipole model by Green's functions. Experimental verification of the viability of the proposed model is provided in Section~\ref{sec:experimental_verification}. The paper concludes with Section~\ref{sec:discussion} where the main contributions of this paper are discussed.


\section{Feedback Manipulation by Dielectrophoresis} 
\label{sec:feedback_manipulation_by_dielectrophoresis}
A great advantage of feedback manipulation by DEP, in contrast to the conventional use of DEP, is that it allows one to manipulate individual microparticles. Nevertheless, this comes with the cost of higher computational demands on the control system because the voltages applied to the electrodes cannot be precomputed anymore and have to be adjusted in real time as required by the feedback loop.

\begin{figure}
\includegraphics[width=8.4cm]{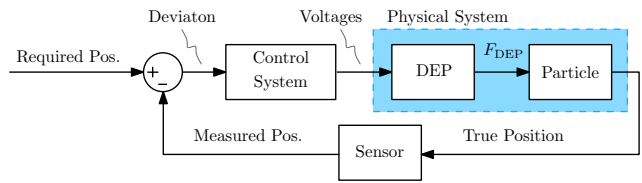}
\caption{\label{fig:feedbackScheme} A block diagram of feedback manipulation by DEP.}
\end{figure}

We can explain this with an aid of Fig.~\ref{fig:feedbackScheme} depicting the scheme of feedback manipulation by DEP. The measured position of the microparticle is subtracted from the required position. The deviation is fed to a control system that calculates a DEP force needed to reduce this deviation---to steer the microparticle towards the required position. Therefore, the voltages on electrodes are required to generate such a force. Such voltages are then applied to the electrodes and the generated DEP force acts on the microparticle moving it towards the required position. The new position is then measured and the whole cycle is repeated. The crucial part of this algorithm is hidden in the control system where, in order to compute the voltages generating the desired DEP force, a model relating the voltages to the DEP force has to be used in real time\cite{Zemanek2015Feedback}.

Nevertheless, exact DEP models are rather complicated to use in real time. For instance, the widely used \textit{dipole model} has the following form: time-averaged dielectrophoretic force acting on a homogeneous spherical particle in a harmonic field is\cite{Hughes2010Nanoelectromechanics}
\begin{eqnarray}
\label{eq:dep}
\nonumber  \langle \mathbf{F}_\mathrm{DEP}(t) \rangle = \pi&&\varepsilon_\mathrm{m}r^3 \Big( \mathrm{Re}\{K(\omega)\} \nabla|\mathbf{E}|^2 + 2\mathrm{Im}\{K(\omega)\}\\ 
  &&\times\left( E_x^2\nabla\varphi_x + E_y^2\nabla\varphi_y + E_z^2\nabla\varphi_z \right) \Big),
\end{eqnarray}
where $\varepsilon_\mathrm{m}$ is the permittivity of the surrounding medium, $r$ is the particle’s radius, $K(\omega)$ is a frequency dependent constant known as Clausius–Mossotti factor, $\mathbf{E}=[E_x,E_y,E_z]$ is the amplitude of the harmonic electric field $\mathrm{Re}\{\mathbf{E}\,\mathrm{e}^{i\omega t}\}$, the phase of the electric field is denoted by $\varphi_a,a\in\{x,y,z\}$ and finally, $\mathrm{Re}\{\cdot\}$ and $\mathrm{Im}\{\cdot\}$ denote real and imaginary parts of a complex number, respectively. For brevity, the spatial dependence is omitted in the notation.

According to~\eqref{eq:dep}, to determine the DEP force due to applied voltages to electrodes, one needs to know the electric field $\mathbf{E}$. The electric field $\mathbf{E}$ is given by $\mathbf{E}=-\nabla\phi$, where the potential $\phi$ is calculated from Laplace equation $\nabla^2 \phi = 0$ with mixed boundary conditions. Orienting the reference frame so that the electrodes lie in the $x$-$y$ plane and manipulated objects are situated above it, the domain is defined by the half-space $z>0$. The boundary conditions are given by the voltages applied to the electrodes (Dirichlet boundary condition) and a zero-flux condition in the normal direction to the electrode plane in the intervening space between the electrodes (Neumann boundary condition)\cite{Morgan2001Dielectrophoretic}. This BVP is analytically intractable and can be solved only approximately by numerical solvers. Since the control algorithm is supposed to run in real time and the calculation of the DEP force must not take more then a few milliseconds, solving on-line this exact BVP numerically is infeasible.

A partial remedy to this issue is to express explicitly the dependence of $\mathbf{F}_\mathrm{DEP}$ on the voltages applied to the electrodes. By superposition, we express the net potential $\phi(x,y,z)$ as a weighted sum of normalized contributions from individual electrodes. That is,
\begin{equation}
\label{eq:superposition}
  \phi(x,y,z) = \sum^n_{i=1} u_i \phi_i(x,y,z), 
\end{equation}
where $n$ is the number of electrodes, $u_i$ serves as a scaling factor given by voltage on $i$th electrode, $\phi_i$ is the contribution to the net potential from the $i$th electrode when $1 \mathrm{V}$ is applied to it while the remaining electrodes are grounded. Now, to determine the net potential $\phi(x,y,z)$, we have to solve $n$ BVPs ($\nabla^2 \phi_i = 0,\,i=1,\dots,n$) that are still analytically intractable, but that do not change with the voltages applied to the electrodes.

One can solve each of these BVPs numerically on a grid of points in advance, store the solution and use it as a look-up table in real time. Nevertheless, this lookup table grows unacceptably large. As an example, let the microparticles be manipulated within an area of size $1500\times1500\times300\,\upmu\mathrm{m}$. If we grid this area equidistantly with points separated by $5\,\upmu\mathrm{m}$, we obtain $300\times300\times60=5,400,000$ points. Na\"{\i}ve implementation of this approach would thus require to store $[E_x, E_y,E_z]$ and their relevant partial derivatives $[\frac{\partial E_x}{\partial x}, \frac{\partial E_x}{\partial y}, \frac{\partial E_x}{\partial z}, \frac{\partial E_y}{\partial y}, \frac{\partial E_y}{\partial z}]$ for each point in order to evaluate $\mathbf{F}_\mathrm{DEP}$ and all that is only for one electrode.

The volume of the data needed to be stored can be reduced by a method introduced by Kharboutly et al.\cite{Kharboutly2009Modeling}. They use a so-called \textit{Boundary Element Method} and instead of storing directly the derivatives of the potential in points spread throughout all the domain, they store precomputed surface charge density in a grid on electrodes. In real time, when it is required to compute the DEP force at a point, the surface charge density is numerically integrated to calculate the electric field and its derivatives at that point and that allows the computation of the DEP force. Nevertheless, for the previous case that still means that a large portion---depending on what extent of the electrode plane is occupied by the electrodes---of $300\times300=90,000$ points have to be stored. Furthermore, the reduction in the volume of data comes at the cost of higher computational complexity because all the stored data points are needed for evaluation of~\eqref{eq:dep}.

In this paper, we propose a different approach. We approximate the previously mentioned, analytically intractable boundary value data so that a closed-form approximated solution can be found.


\section{Green's functions for modeling of dielectrophoresis} 
\label{sec:derivation_and_results}
The solution of the Laplace equation in a half-space domain, which is the case here, with Dirichlet conditions only can be transformed into an integration by use of the Green's theorem. The derivation can be found in Ref. \onlinecite{Wang1996Theoretical} and the resulting formulas are
\begin{equation}
\label{eq:gr_sol_3D}
  \phi(x,y,z) = \frac{z}{2\pi} \iint\limits_{-\infty}^\infty \frac{h(x',y')}{\left[(x-x')^2 +(y-y')^2 + z^2\right]^{3/2}}\,dx'dy'
\end{equation}
for a 3D case and
\begin{equation}
\label{eq:gr_sol_2D}
  \phi(x,z) = \frac{z}{\pi} \int_{-\infty}^{\infty} \frac{h(x')}{(x-x')^2 + z^2}\,dx'
\end{equation}
for the 2D case where one axis, in our case $y$, is redundant, meaning the electrode array has infinitely long electrodes along $y$ axis. The functions $h(x,y)$ and $h(x)$ are Dirichlet boundary conditions that represent values of the potential on the electrode plane, that is $\phi(x,y,0)$ and $\phi(x,0)$, respectively. Thus, to obtain a closed-form description of the potential---and subsequently also of the DEP force---one only needs to compute the integral \eqref{eq:gr_sol_3D} or \eqref{eq:gr_sol_2D}. However, in order to achieve that, it is necessary to know $h(x,y)$ (or $h(x)$) and that means also the potential on the electrode plane in the intermediate space between the electrodes where the mixed boundary conditions impose a zero normal flux. Furthermore, functions $h(x,y)$ or $h(x)$ have to be such that the evaluated integral \eqref{eq:gr_sol_3D} or \eqref{eq:gr_sol_2D} is expressible as a closed-form expression containing only elementary functions; only then is the solution for the potential applicable in real-time feedback control.

To determine the values of the potential on the electrode plane in between the electrodes, various approximations of the decay of the potential away from the electrode can be found in the literature\cite{Morgan2001Dielectrophoretic,Chang2003ClosedForm,Wang1996Theoretical}. Nevertheless, they are all designed only for interdigitated electrode arrays. For more complex electrode array designs, they are either inapplicable or the integrals \eqref{eq:gr_sol_3D} and \eqref{eq:gr_sol_2D} are analytically intractable for the approximation of the potential and have to be solved numerically. Then, however, formulating the solution of the BVP as an integration loses meaning since both the new and the original problem have to be solved numerically.

In this paper, instead, we numerically solve the original BVP with mixed boundary conditions. In order to obtain $h(x,y)$ (or $h(x)$), we approximate this numerical solution on the electrode plane (i.e. on the boundary of the domain) by an analytical model. As we require the integrals \eqref{eq:gr_sol_3D} and \eqref{eq:gr_sol_2D} to be expressible in closed form, we restrict the class of approximating models to piece-wise polynomial models in the 2D case and piece-wise constant models in the 3D case. Having the approximation of the potential on the electrode plane, an approximate closed-form expression for the potential in the half-space domain is obtained by evaluating the integral \eqref{eq:gr_sol_3D} (or \eqref{eq:gr_sol_2D}). Thus, by~\eqref{eq:dep} and~\eqref{eq:superposition}, we also get a model of DEP suitable for feedback control. Note, that the numerical solution is needed only to derive the control-oriented DEP model; the numerical solution is computed only once and off-line. Therefore, all the heavy computational burden is carried off-line and the control system uses the computationally more efficient model in real time. 

It is worth mentioning that since the potential---as the solution of the Laplace equation---is a harmonic function, it is infinitely differentiable\cite{Evans2010Partial}. Furthermore, due to the \textit{Maximum principle}\cite{Evans2010Partial} the error of the approximated potential diminishes as one moves further away from the electrodes. Thus, the accuracy of the model can be controlled by the accuracy of the approximation of the potential on the boundary of the domain.

In the remainder of the paper, we apply the described methodology to two electrode arrays.

\subsection{Example 1: Interdigitated Electrode Array} 
\label{sub:first_example_parallel_electrode_array}
\begin{figure}
\includegraphics[width=8.4cm]{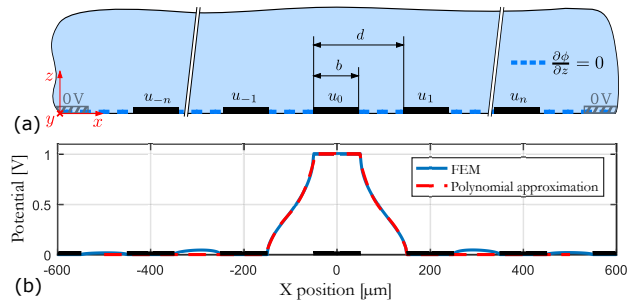}
\caption{\label{fig:example1_scheme} Interdigitated electrode array: (a) a side-view diagram and (b) an approximation of the boundary conditions. The black rectangles represent electrodes and the shaded rectangles on the left and right represent the possibly added grounding plates.}
\end{figure}

Let us consider an electrode array with $(2n+1)$ electrodes, the single electrode width $b$ and center-to-center distance between the electrodes $d$ (see Fig.~\ref{fig:example1_scheme}(a)). We assume that the electrodes are infinitely long and thus drop the dependence on the $y$ coordinate altogether.

We decompose the net potential by~\eqref{eq:superposition} into a weighted sum of normalized contributions $\phi_i$ from individual electrodes. Furthermore, we assume that the potential contribution $\phi_i$ is shift-invariant for a shift $d$ along the $x$-axis. That means the potential contribution $\phi_i$ is identical for each electrode up to a shift. Mathematically,
\begin{equation}
\label{eq:ex1_assumption}
  \phi_i(x,z) = \phi_{i\pm1}(x \pm d, z), \quad i=-n+1,\dots,n-1.
\end{equation}
Clearly, this assumption does not hold for the electrodes close to the perimeter of the electrode array. For instance, the potential contribution $\phi_n (x,z)$ (i.e. from the electrode on the perimeter) decays more quickly towards the $(n-1)$th electrode, which is grounded, than towards the other side, where there is no grounded plate near. Nevertheless, this issue can be resolved (and the assumption~\eqref{eq:ex1_assumption} justified) by manufacturing grounding plates along the perimeter.

As a result of the assumption~\eqref{eq:ex1_assumption}, we have to compute the integral~\eqref{eq:gr_sol_2D} only for one $\phi_i(x,z)$, e.g. $\phi_0 (x,z)$. The remaining potential contributions are determined simply by shifting, that is $\phi_i (x,z) = \phi_0 (x+id,z)$ with $i\in\{-n,\dots, n\}$, and the net potential is then given by~\eqref{eq:superposition}. Nevertheless, to compute~\eqref{eq:gr_sol_2D} for $\phi_0 (x,z)$, we need to know $h(x) := \phi_0 (x,0)$ while the values of $\phi_0 (x,0)$ are known only on the electrodes and unknown on the rest of the bottom boundary. To overcome this problem, we solve the original Laplace equation with mixed boundary conditions numerically by \textit{Finite Element Method (FEM)} in COMSOL Multiphysics. Then, we take the values of the potential on the bottom boundary (i.e. $z=0$) between the $0$-th electrode and its left adjacent electrode and fit a polynomial $p(x)$ to these values. The polynomial approximation of $h(x)$ is
\begin{equation}
  \tilde{h}(x) =
  \begin{cases}
    p(x)  &\quad x\in \big[-d-\frac{b}{2}, -\frac{b}{2}\big),\\
    1 &\quad x\in \big[-\frac{b}{2}, \frac{b}{2}\big],\\
    p(-x) &\quad x\in \big(\frac{b}{2}, d+\frac{b}{2}\big],\\
    0  & \quad \text{otherwise}.\\
  \end{cases}
\end{equation}

Specifically, for an electrode array with parameters $d=2b=200\,\upmu\mathrm{m}$, we fitted a third-order polynomial to the FEM solution and the fitted polynomial is
\begin{equation}
  p(x) = 1.36 \times 10^{-6} x^3 + 4.34 \times 10^{-4} x^2 +  5.15 \times 10^{-2} x +  2.59,
\end{equation}
where $x$ is in micrometers.

The FEM solution together with the polynomial approximation is displayed in Fig.~\ref{fig:example1_scheme}(b). Apparently, the polynomial approximation describes the FEM solution very accurately. However, one should not overlook the small humps in the gaps between the electrodes, which are completely omitted by the approximation. 

\begin{figure}
\includegraphics[width=8.4cm]{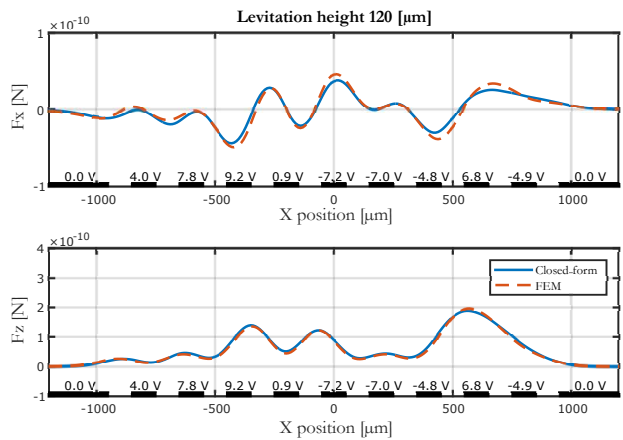}
\caption{\label{fig:example1_comparison} A component-wise comparison of DEP force fields computed for interdigitated electrode array ($d=2b=200\,\upmu\mathrm{m}$) numerically by FEM and analytically based on the approximate closed-form solution for the potential. The force fields are computed for the height of $120\,\upmu\mathrm{m}$ above the electrode array. The applied potentials are indicated above the electrodes represented by the black rectangles. (Multimedia view)}
\end{figure}

With $\tilde{h}(x)$ approximating $\phi_0 (x,0)$ we can compute the integral~\eqref{eq:gr_sol_2D} and obtain an approximate closed-form solution for $\phi_0 (x,z)$. Then, by~\eqref{eq:ex1_assumption} and~\eqref{eq:superposition} we get the net potential $\phi(x,z)$ for any choice of electrode potentials $u_i$. Thus, we can compute the electric field intensity and the pertaining DEP force. We do not present the evaluated integral here, because it is rather lengthy and it would not serve any purpose. Nevertheless, it is crucial to mention that the evaluated integral is indeed expressible as a closed-form expression containing only elementary functions and thus it is easily applicable in real time.

To validate the proposed model, we compare DEP force fields computed by \eqref{eq:dep} for the potential obtained by numerical solution of the original BVP with mixed boundary conditions and for the potential obtained by solution of the approximated BVP. The comparison is shown in Fig.~\ref{fig:example1_comparison} (Multimedia view) . The comparison is carried out for an electrode array with nine electrodes $(n=4)$, with grounded perimeter (also visible in the figure) and with the single electrode width $b=100\,\upmu\mathrm{m}$ and the distance between the electrodes $d=200\,\upmu\mathrm{m}$. The remaining parameters are: $r=25\,\upmu\mathrm{m}$, $\varepsilon_\mathrm{m}=7.08\cdot10^{-10}$ F/m and $K(\omega)=-0.4618-0.1454i$. As usual for the standing wave DEP, the harmonic signal applied on all electrodes has the same frequency and phase. Since it is rather inconvenient to compare the vector fields, the comparison is done for one particular height above the electrodes ($120\,\upmu\mathrm{m}$) and for varying potentials on the electrodes. 

\subsection{Example 2: Four-Leaf Clover Electrode Array} 
\label{sub:second_example_four_sector_electrode_array}
\begin{figure}
\includegraphics[width=8.4cm]{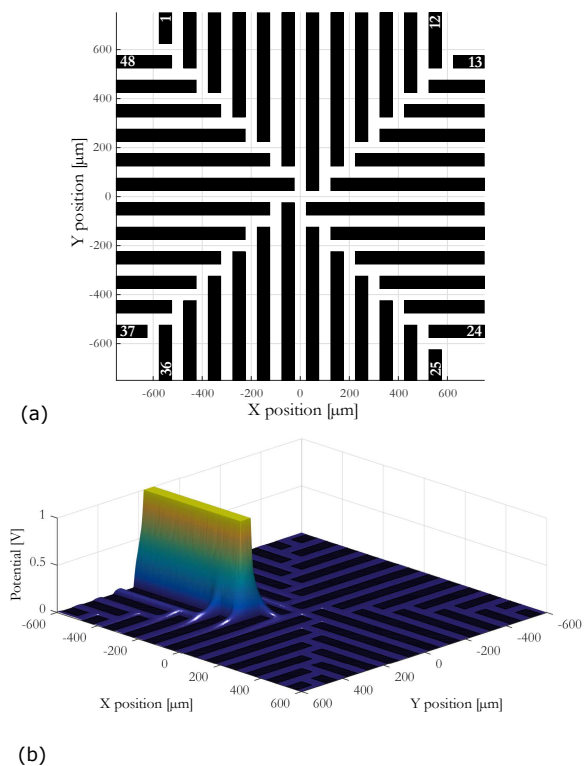}
\caption{\label{fig:example2_scheme} Four-leaf clover electrode array: (a) a top-view diagram and (b) FEM solution for the normalized potential contribution from one electrode.}
\end{figure}

In the second example, we show a related approach how to approximate the values of the potential on the electrode plane for a more complex electrode array shown in Fig.~\ref{fig:example2_scheme}(a). It consists of four quadrants and allows manipulation of microparticles in all three directions above the electrode array, as it was experimentally verified\cite{Zemanek2014Dielectrophoretic}. The width of the electrodes is $b=50\,\upmu\mathrm{m}$ and the center-to-center distance between the electrodes is $d=100\,\upmu\mathrm{m}$. We assume that the electrodes extend to infinity at one end.

Along similar lines as in Example 1, based on the superposition principle~\eqref{eq:superposition} we express the net potential as a weighted sum of normalized contributions $\phi_i$ from individual electrodes and make a similar assumption that the normalized contributions are identical up to a shift and/or rotation with respect to each other; for instance, for electrodes with indexes ranging from 1 to 7 (the indexes are shown in Fig.~\ref{fig:example2_scheme}(a)), it holds that
\begin{equation}
\label{eq:ex2_assumption}
  \phi_{i+1}(x,y,z) = \phi_{i}(x - d, y + d, z),
\end{equation}
where $d$ is the center-to-center distance between the electrodes.

Again, thanks to this assumption, we need to compute the integral~\eqref{eq:gr_sol_3D} only for one $\phi_i (x,y,z)$. In this case, we choose $\phi_{44} (x,y,z)$ because it is close to the center of the sector and, analogously to the previous example, the assumption~\eqref{eq:ex2_assumption} holds best for the electrodes in the center.

\begin{figure*}
\includegraphics[width=.75\textwidth]{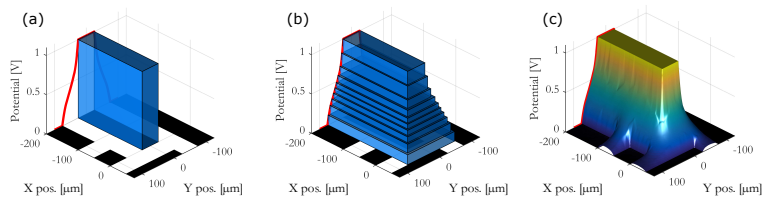}
\caption{\label{fig:example2_approx} An approximation of the desired shape of the bottom boundary condition ($z=0$) for the four-leaf clover electrode array: (a) one block approximation, (b) staircase approximation and (c) boundary condition obtained by FEM. Black rectangles represent electrodes.}
\end{figure*}

It remains to find the approximation of $h(x,y):=\phi_{44} (x,y,0)$. In order to do that, we approximate the FEM solution for $\phi_{44} (x,y,0)$ shown in Fig.~\ref{fig:example2_scheme}(b). This time, however, the integral~\eqref{eq:gr_sol_3D} is analytically intractable for $h(x,y)$ being polynomial or even linear---we are unable to express the integral in closed form for anything other than for constant boundary condition. Thus, instead of using a polynomial or linear approximation, the desired shape of the potential is constructed from blocks. Initially, we approximate the boundary condition in the roughest possible way; we assume that the potential between the electrodes drops immediately to zero as one moves away from the electrode (see Fig.~\ref{fig:example2_approx}(a)). Then summing the ``scaled'' and ``shifted'' versions of this boundary condition (see Fig.~\ref{fig:example2_approx}(b)) approximately gives the desired shape (see Fig.~\ref{fig:example2_approx}(c)).

Let us begin with the one-block approximation. We define the block boundary condition for one-side infinitely long electrode as
\begin{equation}
\label{eq:gr_3d_h0}
  h_0(x,y) =
  \begin{cases}
      1 &\quad  x\leq0 \,\,\, \text{and} \,\,\, y\in \left[-\frac{b}{2}, \frac{b}{2} \right],\\
      0  & \quad \text{otherwise},
    \end{cases}
\end{equation}
where $b$ is the width of the electrode. The staircase approximation of the desired shape is then obtained by
\begin{equation}
\label{eq:gr_3d_h}
  \tilde{h}(x,y) = \sum_{i=1}^N \alpha_i\,h_0 \left(  x-\dfrac{(\beta_i-1)b}{2},  \frac{y}{\beta_i}  \right),
\end{equation}
where $N$ is the number of blocks, $\alpha_i$ determines the height of the block and $\beta_i$ is a scaling parameter, meaning that $\beta_i=2$ scales the block so that it is twice as wide as the original electrode. Notice, that we assumed that the potential decays identically along the $x$ and $y$ axes and thus the coefficients $\beta_i$ determine not only the width but simultaneously also the shift of the blocks along the $x$ axis.

\begin{figure*}
\includegraphics[width=.95\textwidth]{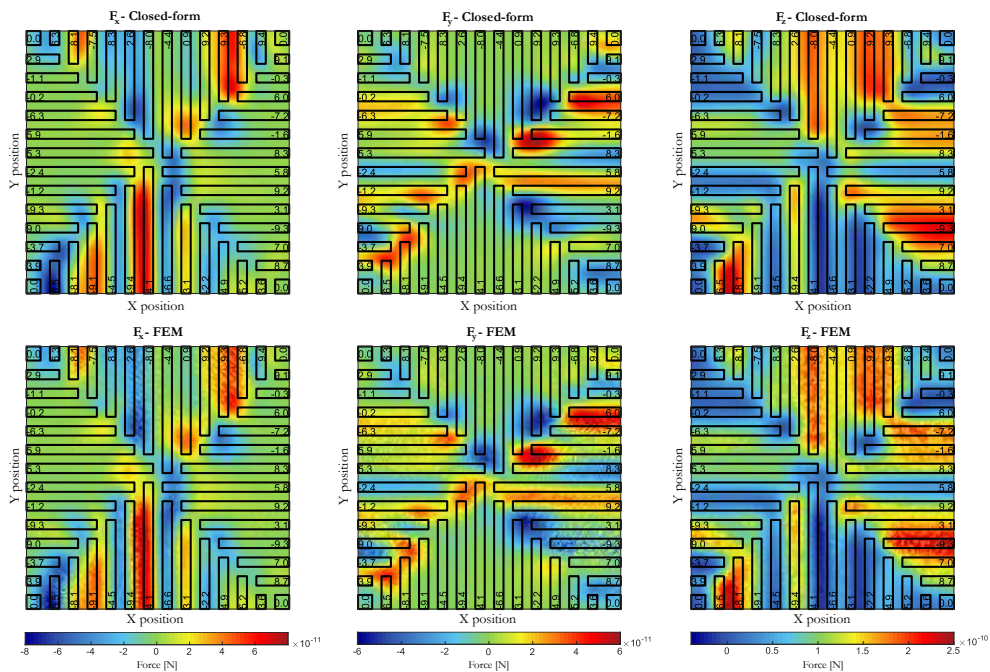}
\caption{\label{fig:example2_comp} A comparison of individual components of DEP force fields calculated for the four-leaf clover electrode array with single electrode width $b=50\,\upmu\mathrm{m}$ and center-to-center distance between the electrodes $d=100\,\upmu\mathrm{m}$. The force fields are calculated numerically by FEM and analytically based on the approximate closed-form solution for the potential. The force fields are computed for the height of $120\,\upmu\mathrm{m}$ above the electrode array. The numerical values inside the electrodes represent applied potentials in volts. (Multimedia view)}
\end{figure*}

Given the FEM solution for mixed boundary conditions, the coefficients $\alpha_i$ and $\beta_i$ in~\eqref{eq:gr_3d_h} can be determined by solving the following optimization problem
\begin{align}
\label{eq:gr_3d_optim}
  \min_{\alpha_i,\beta_i\in\mathbb{R},i=1,\dots,N}  &\,\,\, \| \tilde{h}(x_0,y) - \phi_\mathrm{FEM}(x_0,y,0)\|_2 \\
  \nonumber       \text{subject to:}\,    &\,\,\, \sum_{i=1}^N \alpha_i = 1, \\
  \nonumber                 &\,\,\, \beta_i \in [ 1, 3 ],\quad i=1,\dots,N,
\end{align}
where $\phi_\mathrm{FEM} (x,y,z)$ is the FEM solution. Note that the 2-norm above measures the size of a function of the real $y$ variable, but in the numerical optimization we are only able to consider samples of $y$, which is not encoded in the optimization problem statement for the sake of simplicity. With the assumption that the potential decays identically along $x$ and $y$ axes, both $\alpha_i$ and $\beta_i$ can be determined from a $y$-$z$ cross-section of $\phi_\mathrm{FEM} (x,y,0)$, we fixed $x$ to be a negative constant value $x_0$. For instance, the red curve in Fig.~\ref{fig:example2_approx}(a) represents $\phi_\mathrm{FEM} (x_0,y,0)$ for $x_0=-200\,\upmu\mathrm{m}$. The coefficients $\alpha_i$ have to sum up to one because only then is the height of the piled up blocks equal to one. We assume that $d=2b$ and thus restrict the coefficients $\beta_i$ to the interval $[1,3]$ because then the blocks cannot be narrower than the electrode and they cannot interfere with other electrodes. Even though the optimization task is not convex, it still provides very good results when given a good initial guess. For the initial guess, we let $\beta_i$ grow linearly from $1$ to $3$ and set $\alpha_i$ to be proportional to $\frac{\partial}{\partial y}\phi_\mathrm{FEM} (x_0,y,0)$. Figure~\ref{fig:example2_approx}(b) shows results of the optimization process for $N=10$. 

Having the approximation of the boundary condition $\tilde{h}(x,y)$, one can calculate the integral~\eqref{eq:gr_sol_3D} and obtain a closed-form solution for $\phi_{44} (x,y,z)$. Instead of using the boundary condition $\tilde{h}(x,y)$ composed of several blocks directly, due to linearity of the integral, we can use the one-block boundary condition $h_0 (x,y)$, calculate the integral~\eqref{eq:gr_sol_3D} and compose the closed-form approximation for $\phi_{44}(x,y,z)$ in the same way as $\tilde{h}(x,y)$ is itself composed. This is exactly how we proceed. Substitution of $h_0(x,y)$ into the integral~\eqref{eq:gr_sol_3D} gives
\begin{equation}
  \phi_{i0}(x,y,z)\! =\! \frac{z}{2\pi} \! \int\limits_{-\frac{b}{2}}^{\frac{b}{2}}\!\int\limits_{-\infty}^{0}\!\! \frac{1}{\left((x\!-\!x')^2\!+\!(y\!-\!y')^2\!+\!z^2\right)^{3/2}}\,dx'\!dy'.  
\end{equation}

\begin{figure}[t]
\includegraphics[width=7.4cm]{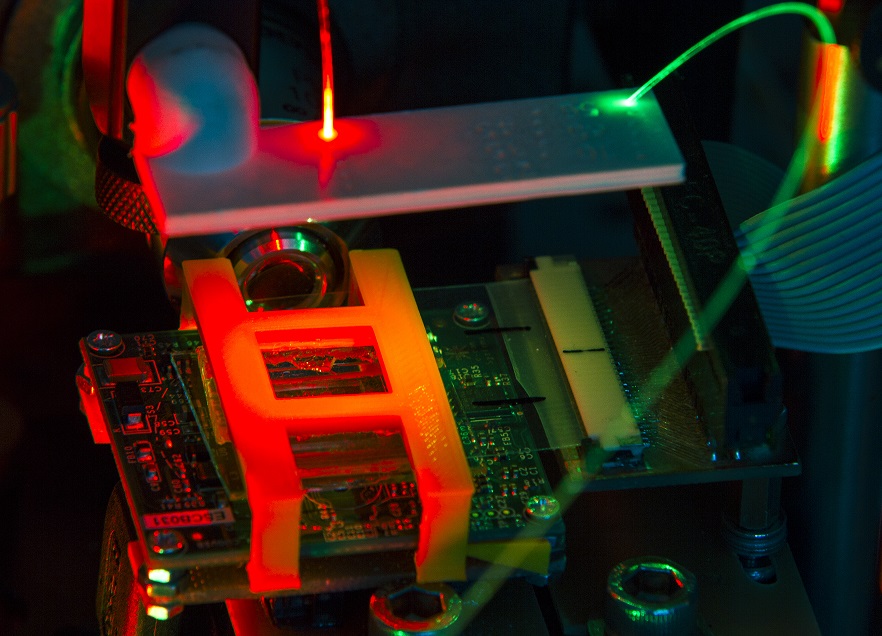}
\caption{\label{fig:expVerification_photo} Photo of the apparatus used for the experimental verification of the proposed control-oriented model of DEP.}
\end{figure}





Again, we do not present the evaluated integral since one can easily compute it in Mathematica, Maple, Matlab or any another computer algebra package. Nevertheless, we emphasize that the evaluated integral is expressible in closed form usable in real time. The approximate closed-form solution for $\phi_{44} (x,y,z)$ is then obtained as
\begin{equation}
  \phi_i(x,y,z) = \sum_{i=1}^N \alpha_i\,\phi_{i0}\left(x-\dfrac{(\beta_i-1)b}{2},\frac{y}{\beta_i},z \right).
\end{equation}

Similarly, as in the 2D case, we do not compare the potentials directly because what we are interested in are the DEP force fields derived from the potentials. Since from the visualization point of view it is rather inconvenient to directly compare 3D force fields, we compare their components separately. The comparisons were carried out for the same parameters as in Example 1 and they are shown in Fig.~\ref{fig:example2_comp} (Multimedia view). Fig.~\ref{fig:example2_comp} shows a comparison carried out for randomly varying potentials on the electrodes. Based on this comparison, the force field computed by the proposed approximate closed-form model is seen to match that computed based on the FEM solution. 


\section{Experimental Verification} 
\label{sec:experimental_verification}
\begin{figure}[t]
\includegraphics[width=8.4cm]{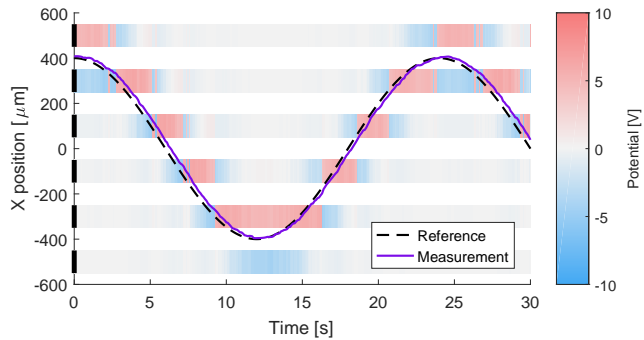}
\caption{\label{fig:expVerification} Experimental verification of the proposed model of DEP in an experiment where a microsphere is steered along a reference trajectory. Only the transverse coordinate (i.e. $x$) is shown. The colors at electrode locations show what potentials were applied to the electrodes at a given time. Notice that $x$ axis here is time and thus any vertical cut shows what potentials were applied to the electrodes at the corresponding time.}
\end{figure}

To verify the applicability of the proposed DEP model, we used it in an experiment where a $50\,\upmu\mathrm{m}$ polystyrene microsphere was manipulated by a control system with a feedback loop. The goal of the control system is to steer the microsphere along a reference trajectory. The microsphere was suspended in water contained in a pool above an interdigitated electrode array with six electrodes and $d=2b=200\,\upmu\mathrm{m}$ (see Fig.~\ref{fig:example1_scheme}(a)). A detailed description of the control and measuring system can be found in Ref.~\onlinecite{Zemanek2015Feedback} and in Ref.~\onlinecite{Gurtner2016TwinBeam}, respectively. Figure~\ref{fig:expVerification_photo} displays a photo of the hardware setup. Figure~\ref{fig:expVerification} shows reference and measured trajectories of the microsphere. In addition, the figure also displays the potentials applied to the electrodes in order to steer the microsphere along the reference trajectory. These potentials were computed in real time by the control system based on the proposed DEP model. It is noteworthy, that in Ref.~\onlinecite{Zemanek2015Feedback} a numerical solution taking approximately $1\,\mathrm{GB}$ of memory space was used to calculate DEP force in real time, whereas here the DEP model is represented by a closed-form analytical expression that takes almost no memory space and is computationally efficient.

\section{Discussion} 
\label{sec:discussion}
\begin{figure}[t]
\includegraphics[width=7.4cm]{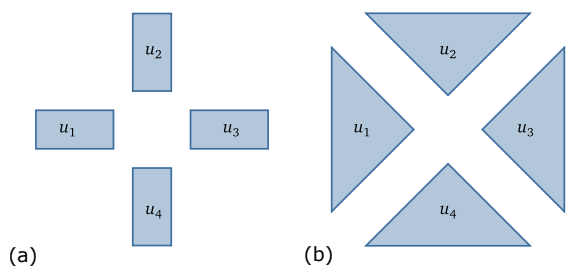}
\caption{\label{fig:discElArrays} Some other electrode array designs in the literature\cite{Kharboutly2013High,Kharboutly2009Modeling} for which the proposed modeling methodology is also applicable.}
\end{figure}

Although the presented approach to modeling of dielectrophoresis is demonstrated on the dipole approximation of the microparticles, it can also be applied to more complex and accurate multipole approximations; no modification would be needed because multipole approximations only require higher derivatives of the potential and our proposed approximation calculates an infinitely differentiable closed-form approximation of the potential. Even though we demonstrate the approach on two concrete electrode array designs, it can also be used for other planar designs exhibiting similar symmetry; two such examples are shown in Fig.~\ref{fig:discElArrays}. Note that full analytical solutions for exact boundary conditions are not possible in such cases.


\section{Conclusion} 
\label{sec:conclusion}
The major benefit of our approximate modeling methodology for a dielectrophoretic force presented here is that in comparison with the standard analytical or FEM-based (numerical) approaches it yields a mathematical model of DEP whose application is feasible in real time (e.g. in cycles of a few milliseconds or so) on a common laboratory PC, and yet the accuracy of the model is sufficient for the purposes of feedback micromanipulation. This methodology combines numerical and analytical models so that the computational burden associated with the calculation of the numerical solution is carried out off-line and based on this numerical solution an approximated closed-form and easy-to-calculate---or briefly control-oriented---model is derived. This approach allows us to derive a control-oriented model for a broader category of electrode array designs than other approaches used in modeling of DEP.


\begin{acknowledgments}
This research was funded by the Czech Science Foundation within the project P206/12/G014 (Centre for advanced bioanalytical technology, http://www.biocentex.cz).
\end{acknowledgments}

\nocite{*}
\bibliography{Remote}

\end{document}